\documentclass[bsl,bibother]{asl}
\usepackage{pstricks}
\usepackage{pst-node}

\def\FOLsubsection#1{\vskip6pt plus0pt minus2pt\par\noindent{\bf {#1}}}

\title[Prospects for mathematical logic]{
The prospects for mathematical logic in the twenty-first century
}
\author {Samuel R. Buss}
\revauthor{Buss, Samuel R.}
\address{Department of Mathematics
\\ University of California, San Diego
\\ La Jolla, CA 92093-0112}
\email{sbuss@ucsd.edu}
\thanks{Supported in part by NSF grant DMS-9803515.}

\author{Alexander S. Kechris}
\revauthor{Kechris, Alexander S.}
\address{Department of Mathematics
\\ California Institute of Technology
\\ Pasadena, CA 91125}
\email{kechris@caltech.edu}
\thanks{Supported in part by NSF grant DMS-9987437.}

\author{Anand Pillay}
\revauthor{Pillay, Anand}
\address{Department of Mathematics
\\ University of Illinois
\\ Urbana-Champaign, UI 61801}
\email{pillay@math.uiuc.edu}
\thanks{Supported in part by NSF grants DMS-9696268 and DMS-0070179.}

\author{Richard A. Shore}
\revauthor{Shore, Richard A.}
\address{Department of Mathematics
\\ Cornell University
\\ Ithaca, NY 14853 }
\email{shore@math.cornell.edu}
\thanks{Supported in part by NSF grant DMS-9802843.}

\def\sectionAuthor#1{ \hfill \\[-10pt] \penalty10000\par\penalty10000 
By #1. \vskip 5pt \penalty10000\par\penalty10000 }

\begin{document}

\begin{abstract}
The four authors present their speculations about the future developments
of mathematical logic in the twenty-first century.  The areas of
recursion theory, proof theory and logic for computer science,
model theory, and set theory are discussed independently.
\end{abstract}

\maketitle

\hspace*{1in}\hfill\parbox{3.1in}{{\em
We can only see a short distance ahead, but we can see plenty
there that needs to be done. }\hfill\\ \hspace*{1ex}\hfill A. Turing, 1950.}\\

\section{Introduction}

The annual meeting of the Association for
Symbolic Logic held in Urbana-Champaign, June 2000,
included a panel discussion
on ``The Prospects for Mathematical Logic in the Twenty-First Century''.
The panel discussions included independent presentations by the four
panel members, followed by approximately one hour of lively discussion
with members of the audience.

The main themes of the discussions concerned
the directions mathematical logic should or could pursue in the 
future.  Some members of the audience strongly felt that
logic needs to find more applications to mathematics; however, there
was disagreement as to what kinds of applications were most likely to
be possible and important.  Many people also felt that applications
to computer science will be of great importance.  On the other hand,
quite a few people, while acknowledging the importance of applications of logic,
felt that the most important progress in logic comes from internal
developments.

It seems safe to presume that the future of mathematical logic
will include a multitude of directions and a blend of these
various elements.  Indeed, it speaks well for the strength of the
field that there are multiple compelling directions for
future progress.   It is to be hoped that logic will be
driven both by internal developments and by external applications,
and that these different directions will complement and strengthen each other.

The present article consists of reports by the four panel members,
at times expanding on their panel presentations.
As in the panel discussion, the presentations are divided
into four subareas of logic.
The topics are ordered as in the panel discussion:
R.~Shore discusses recursion theory in
section~2; S.~Buss discusses proof theory and computer science logic
in section~3; A.~Pillay discusses model theory in section~4; and
A.~Kechris discusses set theory in section~5. 

We, the panel members,
wish to thank Carl Jockusch and the rest of the program
committee for conceiving of the panel topics and inviting us
to participate.

\hfill S. Buss \\

\section{Recursion Theory}
\sectionAuthor{Richard A. Shore}

When I was asked to participate in this panel, the first thing that came to my
mind was the verse from Amos (7:14): I am neither a prophet nor a prophet's
disciple. Nonetheless,
with some trepidation,
I agreed to participate.
When I saw that there was to also be a panel on logic in the Twentieth
Century, I thought that one would have been easier -- after all it has already
happened. In any case, I decided to start, in some sense, with the past.
Rather than a prophet for mathematics, I have in my own past been (somewhat
like Amos) a tender of problems and a gatherer of theorems. To change my ways,
I will start not so much with specific questions and results as with an
attempt to point to attitudes, approaches and ideas that have made recursion
theory what it is and might continue to prove useful in the future. After all,
the true role of the prophet is not to predict the future but to point out the
right path and encourage people to follow it.

For recursion theory, as for logic as a whole, there are great theorems in our
past but perhaps a major share of our contribution to mathematics is in the 
view that we take 
of the (mathematical) world and how it guides us to problems, questions,
techniques, answers and theorems. When I teach the basic mathematical logic
course, I like to say in my first lecture that what distinguishes logic is its
concern for, and study of, the language of mathematics. We study formal
languages, their syntax and semantics and the connections between them. Our
work is motivated by the idea that not only are these and related topics
worthy of study in their own right, as both mathematics and foundations, but
that formalizing and analyzing the language of mathematical systems sheds
light on the mathematics itself.

Along these lines, the language that recursion theory originally formalized
and still studies is that of computability: machines, algorithms, deduction
systems, equation calculi, etc. Its first great contribution, of course, was
the formalization of the notion of a function being computable by an algorithm
and the discovery of many remarkable instances in all branches of mathematics
of the dividing line between computability and noncomputability, decidability
and nondecidability. I see these results as instances of the overarching
concern of recursion theory with notions of complexity at all levels from the
space/time and other subrecursive hierarchies of computer science through
Turing computability and arithmetic to descriptive set theory and analysis
and, finally, to higher recursion theory and set theory. Each way station on
this road of exploration has its appropriate notions of reducibility and
degree structure and its own hierarchies. Our viewpoint is that an analysis of
relative complexity in any of these terms sheds important light on the
fundamental notions of computability and definability and can 
serve to illuminate, distinguish and classify mathematical structures in
useful ways.

I would like to mention a few issues and areas where I think these ideas and
approaches have been useful and I expect will continue to be so in the future.
Of course, there is no expectation of being exhaustive and I admit to
concentrating on those areas that have caught my own interest.

\FOLsubsection{Classical recursion theory.} In general, I view the important
issues in classical recursion theory as analyzing the relations among various
notions of complexity and definability and so also investigating the
possible automorphisms of the computability structures of interest. At the
general level, there are many open problems about the connections between
Turing degrees, rates of growth of functions, set theoretic structural
properties, definability in arithmetic and the jump classes. For a whole array
of specific question in a range of areas, I recommend 
Cholak et al.~\cite{cll2000}
the proceedings volume of the 1999 AMS Boulder conference on open problems in
recursion theory.

The most important and pressing current problem in this area is, I\ believe,
the clarification of the situation with respect to, first, the definability 
of the jump operator and the notion of relative recursive enumerability and, 
second, the 
existence of automorphisms of the Turing and r.e.\ degrees. In the past year,
Cooper's original definition of the jump in the 
Turing degrees~\cite{cooper90,cooper93}
has been shown to be false -- the proposed property does not define
the jump (Shore and Slaman~\cite{ssND}). 
Taking an approach quite different from Cooper's,
Shore and Slaman~\cite{ss99} then proved that the jump is
definable.
Their approach uses results of Slaman and Woodin~\cite{swND}
that strongly employ set
theoretic and metamathematical arguments. While this is pleasing in some ways,
it is unsatisfactory in others. In particular, it does not supply what I would
call a natural order theoretic definition of the jump. For example, the
definition explicitly talks about codings of (models of) arithmetic in the
degrees. (See Shore~\cite{shore99a}.) 

In various versions of Cooper~\cite{cooper00},
Cooper has since proposed two other
candidates for natural definitions of the jump. The first does not define the
jump (Shore and Slaman~\cite{ssND}), and it remains to be seen whether the
second,
much more complicated one, does. Cooper has also proposed a number of ways of
using such a definition of the jump to define relative recursive
enumerability. There are fundamental difficulties to be surmounted in any
attempt to define the notion of r.e.\ by these means (Slaman [personal
communication]), but establishing a natural definition of the jump still seems
to be the most likely route to a definition of recursive enumerability. (See
Slaman~\cite{slaman00}.)

In the area of using classical computability type complexity properties to
classify mathematical structures, I would like to point to the exciting
developments in current work by Nabutovsky and Weinberger~\cite{nwND}
discussed by
Soare in his lecture at this meeting (see Soare~\cite{soareND}). This work uses
complexity properties not just on the decidable/undecidable border but far
beyond. It uses both rates of convergence and r.e.\ Turing degrees to distinguish
interesting classes of spaces and Riemannian metrics by capturing certain
types of invariants in terms of computability properties. We certainly hope
for more such interactions in the future.

\FOLsubsection{Descriptive set theory.}
 Although I am far from an expert, I
am a great fan of descriptive set theory and I view this subject as a major
success for what I have called the recursion theoretic point of view. The
whole array of issues connected to hierarchies, complexity classes,
reducibilities and definability are fundamental to recent work in this area.
It well illustrates the ideas of classification and analysis of mathematical
structures in such terms and the belief that such analysis supplies important
information about the structures. At a deeper level, the ideas of effective
descriptive set theory --- the direct application of notions of computability
for numbers and functions from classical recursion theory to analyze sets of
reals via the relativization of light faced results and other methods ---
permeate many aspects of the area. (See Moschovakis~\cite{ynm1980}
and the forthcoming book of Louveau.)

This topic really belongs to Alekos Kechris and \S 5, but for myself, I also
have hopes for interactions between Borel notions and computability ones in
the areas of effective algebra and model theory. The issue here, as I see it,
is to classify the complexity of mathematical, algebraic and model theoretic
properties in the domain of computable mathematics. This issue arises in many
disguises some of a descriptive set theoretic nature and others more concerned
with computability and relative computability. (Examples here can be found in
Friedman and Stanley~\cite{fs89},
Camerlo and Gao~\cite{cgND}, Hjorth~\cite{hjorth99}, White~\cite{white00}
and Hirschfeldt, Khoussainov, Shore and Slinko~\cite{hkssND}.)

\FOLsubsection{Effective and Reverse Mathematics.} Perhaps, with the revival
of interest in computational approaches to much of mathematics, we will see
more interest in some of the topics of effective mathematics such as
determining what input information about structures is needed to compute
various properties, functions and so on. The technical notions involved here
include versions of intrinsic computability, degree spectra and the like. (For
these specific topics see, for example, Khoussainov and Shore~\cite{ks99}
or Shore~\cite{shore99b}
and, for the whole range of issues involving effective mathematics, the
handbook Ershov et al.~\cite{egnr98}.)

These issues in effective mathematics are also related to the foundational
concerns of reverse mathematics which, on its face, uses another, proof
theoretic, yardstick to measure complexity. The proof theoretic measures used,
however, turn out to be intimately connected to ones studied in recursion
theory from relative computability and the Turing jump to relative
hyperarithmeticity and the hyperjump. The basic reference here is 
Simpson~\cite{simpson99}.
The foundational issues addressed by reverse mathematics are important
ones and we expect the contributions and approaches of recursion theory in
this area to increase and prove important as well. A current snapshot of
work in the field is provided by the collection Simpson~\cite{simpsonND}.

Along with the rise of mathematical interest in computational procedures come
other views of computability that should be worth investigating. These should
be measures that reflect mathematical practice in particular domains rather
than just what we traditionally view as computation in our traditional
discrete, digital approach. One obvious candidate is the notion of computation
introduced by Blum, Shub and Smale~\cite{bss99}
(see also Blum et al.~\cite{bcss97}). Here
the corresponding notions of complexity have been used (for example, in 
Chong~\cite{chong94,chong95})
to distinguish interesting phenomena about Julia sets and the
like in dynamics. I would suggest that we should look for other applications
and other notions of computability appropriate to various mathematical domains
as well.

\FOLsubsection{Set Theory.} Next, I would like to mention (general) set
theory as an area for uses of the recursion theoretic world view. I view the
early development by Jensen of the fine structure of L~\cite{jensen72}
as another
outstanding example of the application of the this viewpoint
to open up a whole line of analysis and investigation. Its extensions continue
to grow remarkably on their own as witnessed by the talks of Steel and Neeman
at this meeting and, for example, L\"{o}we and Steel~\cite{ls1999}.
There is also
still room for applications of definability, complexity and effective analysis
in the setting of classical set theoretic problems as witnessed, for example,
by recent work by Slaman~\cite{slaman99} and Groszek and Slaman~\cite{sg99}.

Saving the best for last, we come to computer science.

\FOLsubsection{Computer Science.} The origins of logic were in foundations
and philosophy; most of us on this panel, and in the audience, were trained and
grew up as mathematicians; the primary future growth opportunities of logic,
however, clearly lie in computer science. I would make the analogy that logic
is to computer science what mathematics is to physics, and vice versa. Logic
serves as the language and foundation for computer science as mathematics does
for physics. Conversely, computer science supplies major problems and new
arenas for logical analysis as physics does for mathematics. This relationship
has affected, and will affect, not just recursion theory but all of logic. There
are far too many instances to even mention but I point to a symposium last
year at the American Association for the Advancement of Science titled ``On
the unusual effectiveness of logic in computer science'' reported on in
Halpern et al.~\cite{hhkvvND}
 as one multifaceted indicator of the view from computer science.

Of course, Sam Buss has much more to say about logic and computer science in
\S 3, but for now I'll put in a word from my own sponsor, recursion theory. The
notions of reducibility, complexity measures, and hierarchies are fundamental
to theoretical computer science. Both below and above the level of polynomial
time computability we have analogs of the classical reducibilities including
one-one, many-one, (bounded) truth-table, weak truth table and Turing that
reflect various views of boundedness of different resources. We also see
analogs of notions from higher recursion theory such as fixed points,
inductive definability and admissible ordinals. Interestingly, new notions 
have been developed at low complexity levels that present alternate notions of
computability involving nondeterminism and probabilistic procedures among
others. The relationships among these various restricted (and extended)
notions of computation form the core questions of complexity theory in
computer science. What are the relationships among the classes $P,NP,PP,BPP$
and so on. We hope that the methods and insights of recursion theory will play
a role in the solution of these fundamental problems of computability.

These concerns will continue to illuminate the investigations of computation
from both practical and theoretical vantage points. More radical innovations
are needed, however, to better reflect the finiteness of all types of 
resources. This is a challenge for computer science, computability theory and
logic as a whole. Linear logic and finite model theory represent some attempts
at addressing these issues, but others are needed that incorporate finiteness
and boundedness of resources in other ways. Nonetheless, our view from
infinity will continue to lend perspective (as in the role of uniformity) and
so play a role in analyzing the finite as well.

In a different direction, we return to the original language of computation. 
Here the beginnings of recursion theory have already played an important role, 
e.g. the Turing machine model as a basic one for computation and the 
$\lambda$-calculus as one for programming languages both abstract and actual. 
And so we come back to the beginnings of the study of the formal languages of 
computation. Along these lines, I would like to close with three, certainly not 
original and probably pie-in-the-sky, problems.

1. ``Prove'' the Church-Turing thesis by finding intuitively obvious or at
least clearly acceptable properties of computation that suffice to guarantee
that any function so computed is recursive. Turing~\cite{turing36}
argues for the
thesis that any function that can be calculated by an abstract human being
using various mechanical aids can be computed by a Turing machine (and so is
recursive). Gandy~\cite{gandy80} argues that
any function that can be calculated by a
machine is also Turing computable. Deutsch~\cite{deu85}
approaches this issue from a
more quantum mechanical perspective. Martin Davis has pointed out (personal
communication) that one can easily prove that computations as given by
deductions in first order logic relations from a finite set of sentences about
numerals and the function being defined are necessarily recursive. An analysis
based on the view that what is to be captured is human mechanical
computability is given in Sieg~\cite{sieg97}. 

Perhaps the question is whether we can
be sufficiently precise about what we mean by computation without reference to
the method of carrying out the computation so as to give a more general or
more convincing argument independent of the physical or logical
implementation. For example, do we reject the nonrecursive solutions to
certain differential equations as counterexamples on the basis of our
understanding of physics or of computability. Along these lines, we also
suggest two related questions.

2. What does physics have to say about computability (and provability or
logic)? Do physical restrictions on the one hand, or quantum computing on the
other, mean that we should modify our understanding of computability or at
least study other notions? Consider Deutsch's~\cite{deu85}
Church-Turing principle
and arguments that all physically possible computations can be done by a
quantum computer analog of the universal Turing machine. He argues, in
addition, that the functions computable (in a probabilistic sense) by a
quantum Turing machine are the same as the ones computable by an ordinary
Turing machine, but that there is, in principle, an exponential speed-up in the
computations. How do these considerations affect our notions of both
computability and provability? For some of the issues here see 
Deutsch et al.~\cite{delND}.

3. Find, and argue conclusively for, a formal definition of algorithm and the
appropriate analog of the Church-Turing thesis. Here we want to capture the
intuitive notion that, for example, two particular programs in perhaps
different languages express the same algorithm, while other ones that compute
the same function represent different algorithms for the function. Thus we
want a definition that will up to some precise equivalence relation capture
the notion that two algorithms are the same as opposed to just computing the
same function. Moschovakis~\cite{ynm98} is an interesting approach to this
problem from the viewpoint that recursion, and an appropriate formal
language for it, should be taken as basic to this endeavor.

\section{Proof Theory and Logic for Computer Science}
\sectionAuthor{Sam Buss}
\def\BussTableOne{1}
\def\BussTableTwo{2}

I discuss in this section prospects both for proof theory and for
computer science logic.  I will first
present a broad overview of the presently
active areas in proof theory and computer science logic and
my opinions on which areas are likely to be important in the
future.  I then make some specific predictions about future
developments in these areas.  The section concludes with
an exhortation for mathematical logicians to pay more attention to applications
of logic in computer science.

I am charged with the task of discussing both proof theory and
computer science logic in this section.  It thus has happened
that proof theory is somewhat shortchanged.  The reader who wishes
to seek more comments on proof theory can find 
opinions by a large group of proof theorists in the compendium
prepared by S.~Feferman~\cite{Feferman:proofsurvey}.

\FOLsubsection{Proof theory.}  
I first present a very quick overview of the present goals of 
proof theory. 
Table~\BussTableOne{} gives a ``three-fold''
view of proof theory,  in which proof theory is split into three
broad categories based on the goals of the work in proof theory.

\begin{figure}[t]
\begin{center}
{\large Three-fold View of Proof Theory}
\end{center}
\def\tbwidth{1.4in}  
\newenvironment{tablelist}{%
\begin{list}{}{\itemsep0pt \parsep0pt \topsep1pt \itemindent=-\parindent %
\listparindent=\parindent \leftmargin=\parindent \labelwidth=-\labelsep}}%
{\end{list}}
\begin{tabular}{ccc}
\hline \hline
Stong Systems & Weak Systems & Applied Proof Theory \\
\hline \hline
\parbox[t]{\tbwidth}{\raggedright
\vskip 4pt
\noindent\underline{\em Topics:}
\begin{tablelist}
\item[-] Central foundations of proof theory.
\item[-] Ordinal analysis.
\item[-] Fragments of type theory and set theory.
\end{tablelist}
\par
\vskip 4pt
\noindent\underline{\em Goals/directions:}
\begin{tablelist}
\item[-] Constructive analysis of second-order logic and stronger theories.
\item[-] Methods and tools used extensively in the other areas.
\end{tablelist}
\vspace*{4pt}
}
&
\parbox[t]{\tbwidth}{\raggedright
\vskip 4pt
\noindent\underline{\em Topics:}
\begin{tablelist}
\item[-] Expressive but weak systems, including bounded arithmetic.
\item[-] Complexity (esp., low-level).
\item[-] Proof complexity.
\item[-] Propositional logic.
\end{tablelist}
\par
\vskip 8pt
\noindent\underline{\em Goals/directions:}
\begin{tablelist}
\item[-] Central problem is P vs.\ NP and related questions.
\end{tablelist}
}
&
\parbox[t]{\tbwidth}{\raggedright
\vskip 4pt
\noindent{\em \underline{Topics}:}
\begin{tablelist}
\item[-] Very diverse
\item[-] Theorem proving.
\item[-] Logic programming.
\item[-] Language design.
\item[-] Includes logics which are inherently not first-order.
\end{tablelist}
\par
\vskip 16pt
\noindent\underline{\em Goals/directions:}
\begin{tablelist}
\item[-] Again: very diverse.
\item[-] Central problem is the ``AI'' problem of developing
``true'' artificial intelligence.
\end{tablelist}
}
\\
\hline \hline
\end{tabular}
\vskip4pt
\noindent
Table~\BussTableOne{}.  The areas of proof theory,
organized by goals.
\end{figure}

The first column represents the traditional,
classic approaches to mathematical
proof theory: in this area the goal has been to
understand stronger and stronger systems, from second-order logic up through
higher set theories, and especially to give constructive analyses of
the proof-theoretic strengths of strong systems.  This part
of proof theory has of course seen outstanding progress in the past century,
but in recent times has had more limited success.  The work on constructive
analyses of strong systems has become stymied by technical 
difficulties, and progress has tended to be incremental.  Further
significant progress will require a substantial breakthrough in applicability,
say to all of higher-order logic, as well as a breakthrough
in technical simplicity.  Lest the assessment of this area of proof theory
seem too harsh, I hasten to add that the methods and results of this area
are fundamental to the other areas of proof theory.

The second column represents the smallest area of proof theory.
The presence
of this category is justified by the fact I am presenting the
areas of proof theory categorized by their goals.  The essential goal of this
area is the resolution of the important questions in computational
complexity such as the P versus NP question.  This question turns out to
be very closely linked to corresponding questions about provability, about
proof complexity, and about proof search in very weak proof systems,
including systems as weak as propositional logic.  It is an amazing fact
that very fundamental and simple questions about propositional logic
are still open!  I will make predictions about the future of these
questions below.

The third column represents the broadest branch of proof theory; it is
also the oldest in that it predates the modern mathematical development
of proof theory, with its essential goals stated already by Leibniz.
One goal of this area is to provide logical systems strong enough to
encompass more and more of human reasoning.
A second goal is the development of true AI, 
or ``artificial intelligence''.  Of course, this area of proof theory is
extremely diverse, and it includes
the  aspects of proof theory that deal with
reasoning in restricted domains and aspects of proof theory that are
applicable to programming languages, etc.

I wish to avoid philosophical issues about
consciousness, self-awareness and what it means to have a soul, etc., and
instead seek a purely operational approach to artificial
intelligence.  Thus,
I define artificial intelligence as being constructed systems
which can reason and interact both syntactically and semantically.  
To stress the
last word in the last sentence, I mean
that a true artificial intelligence system
should be able to take the meaning of statements into account, or at least
act as if it takes the meaning into account.
There is some debate about whether logic is really a possible foundation
for artificial intelligence.  The idea that logic should be the 
foundation for AI has fallen out of favor;
indeed, much of the work of artificial
intelligence today is done with non-discrete systems such as neural nets,
which would not count as part of proof theory.  To the
best of my knowledge, there is only
one large-scale present-day attempt to build an AI system based on
logic, namely the {\sc Cyc} system, and this so far has not reported
significant success in spite of a massive effort.  Nonetheless, it is
my opinion that purely analog systems such as neural nets will not
provide a complete solution of the AI problem; but rather,
that discrete processing, including proof theoretic aspects, will be
needed for constructing AI systems.

Of course, most present day work in applied proof theory is not
aimed directly at the AI problem.  There is a large amount
of work being done to extend logic beyond the domain of first-order logic.
This includes, for instance, non-monotonic logics, modal and dynamic logics,
database logics, fuzzy logic, etc.  What these rather disparate areas 
have in common is
that they all wish to extend logic well beyond the
boundaries of the kind of first-order logic that has been successful 
in the foundations of mathematics.

I make some specific predictions about the prospects for
artificial intelligence in a later section.

\FOLsubsection{Logic for computer science.}
A skeletal overview of the present state of affairs for
applications of logic to computer science is presented 
in Table~\BussTableTwo{}, which is titled the
``octopus of logic for computer science.''
I have not attempted to make a definitive summary of the applications
of logic to computer science in Table~\BussTableTwo{},
as they are far too numerous and varied for me to make such an attempt.
The main point of the table is to illustrate
how diverse and extensive the
applications of logic for computer science have become.

Computer science has strong interactions with most of the
traditional areas of logic, with the sole exception of
set theory.  First, the early developments
of both the theory and practice of computer science were very
closely linked to the development of recursion theory, beginning with the
emergence of the stored program paradigm arising from Turing's
model for universal computers.  In more recent times,
recursion theory has been less closely linked to computer
science; however, developments in
complexity theory have often been inspired by constructions in
recursion theory.   Proof theory has many connections to computer science;
indeed, the bulk of the work in the third category of `applied proof
theory' in Table~\BussTableOne{} is oriented towards applications in
computer science.  In addition, the second category of proof theory has
been found to have many connections to complexity theory.  There are likewise
many applications of model theory in computer science as well: in
Table~\BussTableTwo{} these include finite model theory, database theory,
and model-checking.  More generally, whenever one deals with the semantics
of a language, one is implicitly doing model theory.

\def\tbwidthTwo{1.65in}
\def\tbwidthCenter{1.0in}

\begin{figure}[t]
\begin{pspicture}(0,0)(13,18.6)
\rput(6.5,9.7){%
\ovalnode{Z}{%
\parbox{\tbwidthCenter}{\center{\Large\bfseries\em Logic for\\Computer\\Science}}
}
}

\rput[tl](10,11.7){%
\rnode{A}{%
\parbox[t]{\tbwidthTwo}{\raggedright
\underline{\em Complexity theory} \\
Reducibility \\
Oracles \\
Feasible complexity \\
P vs.\ NP \\
Circuit complexity \\
Parallel complexity \\
Finite model theory \\
Diagonalization \\
Natural Proofs \\
Proof complexity \\
Craig interpolation \\
Learning theory \\
Bounded arithmetic
}
}
}
\nccurve[angleB=180,angleA=10,offsetB=2,nodesepB=0.1]{-}{Z}{A}
\rput[tl](9,4.3){%
\rnode{B}{%
\parbox[t]{\tbwidthTwo}{\raggedright
\underline{\em Probabilistic computation} \\
Randomized \\ \hspace*{10pt}computation \\
Probabilistic proofs \\
Interactive proofs \\
PCP, Holographic proofs \\
Quantum computing
}
}
}
\nccurve[angleA=-45,angleB=145,nodesepB=0.1]{-}{Z}{B}
\rput[tl](4.5,3.2){%
\rnode{C}{%
\parbox[t]{\tbwidthTwo} {\raggedright
\underline{\em Verification} \\
Program correctness \\
Hardware verification \\
Fault-tolerance \\
Proof-carrying code \\
Liveness/safeness
}
}
}
\nccurve[angleA=270,angleB=90,offsetB=-1.5,nodesepB=0.2]{-}{Z}{C}
\rput[tl](0.5,5.2){%
\rnode{D}{%
\parbox[t]{\tbwidthTwo} {\raggedright
\underline{\em Language design} \\
Programming languages \\
Denotational semantics \\
Query languages \\
Grammars/parsing \\
Automata theory \\
Natural language \\ \hspace*{10pt}processing
}
}
}
\nccurve[angleA=235,angleB=45,offsetB=-0.7,nodesepB=-0.3]{-}{Z}{D}
\rput[tl](0,10.2){%
\rnode{E}{%
\parbox[t]{\tbwidthTwo} {\raggedright
\underline{\em Strong proof systems} \\
Polymorphism \\
Object-oriented \\ \hspace*{10pt}languages \\
Abstract datatypes \\
$\lambda$-calculi \\
Combinatory logics \\
Functional programming \\
Category theory \\
Realizability
}
}
}
\nccurve[angleA=170,angleB=55,offsetB=-1,nodesepB=-0.3]{-}{Z}{E}
\rput[tl](0.5,16.4){%
\rnode{F}{%
\parbox[t]{\tbwidthTwo} {\raggedright
\underline{\em Weak proof systems} \\
Resolution \\
Logic programming \\
Constraint logic \\ \hspace*{10pt}programming \\
Theorem provers \\
Equational logics \\
Term rewriting \\
Behavioral logics \\
Nonmonotonic logics \\
AI \\
Model checking
}
}
}
\nccurve[angleA=135,angleB=-15,nodesepB=-0.5]{-}{Z}{F}
\rput[tl](5,17.6){%
\rnode{G}{%
\parbox[t]{\tbwidthTwo} {\raggedright
\underline{\em Real computation} \\
Real closed fields \\
Geometry \\
Complexity of \\ \hspace*{10pt}real computation \\
Hybrid systems \\
Computer algebra \\ \hspace*{10pt}systems
}
}
}
\nccurve[angleA=90,angleB=240,offsetB=1.3,nodesepB=0.3]{-}{Z}{G}
\rput[tl](9,16.6){%
\rnode{H}{%
\parbox[t]{\tbwidthTwo} {\raggedright
\underline{\em Other logics} \\
Database languages \\
Least fixed points \\
Modal logics \\
Dynamic logics \\
Theories of knowledge \\
Resource-aware logics \\
Linear logic
}
}
}
\nccurve[angleA=45,angleB=190,nodesepB=0.1,offsetB=0.3]{-}{Z}{H}
\end{pspicture}
\begin{center}
Table \BussTableTwo{}. The Octopus of Logic for Computer Science.
\end{center}
\end{figure}

\FOLsubsection{Future directions and some predictions.}
The above overviews of proof theory and of logic for computer science
indicate the directions that I feel will be their most important areas
for future development.  To be even more explicit  about my expectations
for the future, I will now make a series of quite specific predictions
about when we may solve the important problems in these areas.
\medskip

{\em P versus NP and related questions.}  Although progress in actually
solving the P versus NP problem has been slow, there has been a vast
amount of work related to P, NP and other complexity classes.  I do not
believe that there should be any inherent reason why a solution to the
P versus NP problem should be difficult, but rather think we just need
to find the right idea.  Thus I make the following prediction.
\vspace*{1em}
\vskip0em plus 0.2em minus 0.2em

\noindent
{\bf Prediction 1.}  {\em The P versus NP problem (and many related 
questions in complexity theory) will be solved by the following
date:\footnote{These predictions were formulated in June 2000.}
\begin{center}
2010  \quad $\pm 10$ years.
\end{center}
}

I further predict that P is distinct from NP; however, I am agnostic
about the truth of many of the commonly conjectured cryptographic
conjectures.   

I {\em hope} that the solution to the P versus NP problem will be
some kind of extension of the diagonalization method, that is to say,
that there will be some logical reasoning extending ideas of
self-reference, which will be able to resolve the P/NP problem.
The alternative to a logical solution of this type would be a combinatorial
proof more in the lines of the so-far obtained circuit lower bounds
of Yao, Hastad, Razborov, Smolensky, and others.  To my mind, a combinatorial
proof would be a bit disappointing and a logical proof would be far
preferable.  Obviously, a logical proof would be a tremendous boost
to the prestige and importance of logic.

Two promising recent approaches to solving the P versus NP problem include
recent work on diagonalization (by Fortnow and others) and on
natural proofs and Craig interpolation (beginning with the work of
Razborov and Rudich).  More broadly, much work in weak first-order theories
and on propositional proof complexity is motivated by the desire to find
a logical proof that $\hbox{P}\not=\hbox{NP}$.

\medskip

{\em Future problems in proof complexity.}  Probably the most important
problem in proof complexity is to better understand the structure of
propositional and first-order proofs with cuts.

Another very important problem is to either find, or prove impossible,
proof search procedures which both (a)~are more efficient than
human mathematicians, and (b)~yield humanly intelligible proofs.
However, I think this problem is extremely difficult and can be
accomplished only with the solution of the next problem.

\medskip

{\em Artificial intelligence.}  As discussed above, true AI will
involve semantic reasoning based on machine ``understanding.''  I
do not expect that artificial intelligence will be an all-or-nothing
event of the kind frequently envisioned in popular literature where
we one day suddenly discover that machines have become intelligent.  I also
think that some of the currently expressed fears about the dangers
of artificial intelligence are way over-blown.  Rather, I predict that
progress in artificial intelligence will be a long, slow process of
incremental gains.  Nonetheless, I make the following prediction.

\vspace*{1em}
\vskip 0em plus 0.2em minus 0.2em

\vbox{
\noindent
{\bf Prediction 2.}
{\em
There will be limited but significant success in artificial intelligence
by
\begin{center}
 2050 \quad $\pm 30$ years.
\end{center}
}
}

As discussed earlier, I predict that success in artificial intelligence
will require logic-based reasoning.
By ``limited, but significant success'', I envision that artificial
intelligence may be successful in some relatively broad domain of
knowledge which is generally acknowledged as involving 
operational understanding of semantic concepts.

One good possibility
for a first knowledge domain for the initial artificial intelligence
systems is the area of mathematical reasoning.  There are several
advantages to mathematical reasoning as a knowledge domain.  Firstly,
a computer can interact more-or-less on an equal footing with a human
since no physical interaction is required.  Secondly, the domain is
precisely describable with fixed rules.  Thirdly, reasoning in mathematics
requires both creativity and
a significant semantic understanding of the subject matter,
and thus represents a significant challenge for an AI system.

This leads to the next prediction.

\vspace*{1em}
\vskip 0em plus 0.2em minus 0.2em

\vbox{
\noindent
{\bf Prediction 3.}
{\em
Computer databases of mathematical knowledge will contain,
organize, and retrieve most of the known mathematical literature,
by 
\begin{center}
 2030 \quad $\pm 10$ years.
\end{center}
}
}

\noindent
The first step in fulfilling this prediction is to design a formal
language which can faithfully represent mathematical objects
and constructions in a flexible, extensible way.  Perhaps an
object-oriented language would be a good choice for this; however,
present-day object-oriented languages are not adequate for
representing mathematical objects.

One of the original stated goals of mathematical logic was to provide
a foundational understanding of mathematics.  Quite possibly, the next
major step forward in the foundations of mathematics will occur in
conjunction with the
development of systems fulfilling Prediction~3 or perhaps even with AI
systems for mathematical reasoning.

\FOLsubsection{The relation of logic and computer science.}
As illustrated in the ``octopus'', the area of logic for computer 
science is a very active, vital and diverse discipline.  Indeed, it is
likely that there are more people working on logic within computer
science than outside of computer science.  

In addition, many of the recent developments in computer science
call into question the fundamental concepts of mathematical
logic.  For instance, the introduction of probabilistic proofs and
interactive proofs and the possibility of quantum computing, threaten
the correctness of two of the most fundamental notions in logic,
namely the notions of ``proof'' and ``computability''.\footnote{%
In his talk at the Annual ASL Meeting in June 2000, A.~Widgerson
gave an illuminating survey of some of these new notions for proof
and computability that have arisen from the study of the mathematical
foundations of cryptography using notions
of probabilistic proof and interactive proofs and based on complexity
conjectures.}

However, the so-called core areas of logic have historically slighted
or ignored developments in computer science.  Of course, this is
not universally true and there are numerous examples of cross-over
research; furthermore, in recent years, the use of logic in computer
science has reached a critical mass and it is no longer really possible
for core areas of logic to ignore applications of logic for
computer science.  Nonetheless, I think most people would agree that there is
a significant cultural separation between the traditional areas of logic
and the use of logic for computer science.   

This separation started before my time, so it is difficult for
me to say with any certainty why it occurred. 
But my impression is that the
separation arose in part because, as the field of theoretical
computer science began, the work lacked focus, seemed somewhat
ad hoc and overly concrete, and sometimes lacked depth.
(Of course, this is not unexpected in
a field which was still in its formative stages.)
By comparison, logic in the 1960's was embarked on
a grand project of building coherent and deep theories about
large-scale concepts, such as large cardinals, higher notions
of computability, stronger constructive theories, etc.  The work
in computer science took the opposite direction of looking at
low-level complexity, expressibility and provability in weak
languages, etc.  After about fifty years of work, theoretical
computer science has reached the point where on one hand it is
a mature field with deep and far-reaching results, but, on the other hand,
still has extremely basic open questions: questions such
as whether P is equal to NP or whether mathematically
secure cryptography is possible.

Stronger ties between mathematical logic and computer science certainly
need to be encouraged.
The field of theoretical computer science has grown extremely
large, but is still very much in a formative stage, with many key
theorems still to be proved and very many advances
still needed.  Theoretical computer science offers many new
applications of logic, and challenges or extends
many of the fundamental notions of mathematical logic.

{\em Acknowledgement.} I wish to thank Jeff Remmel for comments on
an earlier draft of this section.

\section{Model Theory}
\sectionAuthor{Anand Pillay}

My aim is to describe some trends and perspectives in model
theory. This article is an expanded version of my talk in
the ``Panel of the Future" at the ASL meeting in Urbana. I
will also incorporate some points which came out during the
subsequent discussion, and so this article may have a
somewhat polemical flavour.

Wilfrid Hodges' book {\em Model Theory}  \cite{Hodges} is a
basic text for the subject and its comprehensive bibliography
can be used as a reference for much of the work
cited in the present article, in particular for everything
in the introduction.

As in Kechris' article it is useful to distinguish between
internal and external aspects of research in model theory.
One could also call these aspects inward and outward-looking.
Of course this distinction is not clear-cut,
and in fact I want to describe a remarkable
unification that has been in process for a few years. In any
case, this inward versus outward dichotomy in no way
corresponds to logic (or foundations) versus mathematics (or
applications).

By inward-looking I mean the development and study of
concepts, problems, etc., proper to model theory itself.
Included here are the compactness theorem for first
order logic (G\"odel, Malcev, Tarski), the theory of
quantifier-elimination and model completeness (Tarski,
Robinson), homogeneous-universal and saturated models
(Morley-Vaught), countable models of complete theories
(Vaught), omitting types, products (Feferman-Vaught),
generalized quantifiers, infinitary logics,... Shelah's work
on classification theory, following Morley's work on
uncountably categorical theories, possibly represented the
first fully-fledged {\em program} within model theory
proper.  The nature of the problem, as well as various
theorems of Shelah himself, allowed him to restrict his
attention to a rather small class of first order theories,
the {\em stable} ones, for which a deep theory was developed.

By outward-looking I
mean the use of model-theoretic methods in the study of
specific structures or theories from mathematics and even
from logic itself. Early such work was Malcev's
use of the compactness theorem to prove local
theorems in group theory. One should include
completeness, model-completeness, and quantifier-elimination
results for abelian groups (Szmielew) and various classes of
fields such as real-closed (Tarski), algebraically closed
(Robinson, Tarski),
$p$-adically closed (Ax-Kochen, Ershov, Macintyre),
differentially closed (Robinson) etc., and resulting
applications. One should also include here nonstandard
analysis as well as the use of model-theoretic methods
(such as nonstandard models) in set theory and in the study
of Peano arithmetic and its fragments. Among the past
successes of ``outward-looking" model theory are the
Ax-Kochen-Ershov analysis of Henselian valued fields and the
resulting asymptotic solution to a conjecture of E. Artin.

\FOLsubsection{The current situation.}
The 70's and 80's saw something of a separation between (i)
those interested primarily in model theory as a tool for
doing mathematics (or logic) and (ii) those, often working
in and around stability theory, for whom model
theory was also an end in itself. This separation is again
only an approximation to the truth: there were people on both
sides (and also on neither side), and already results of
Zilber,  Cherlin, Harrington and Lachlan, and Macintyre, had
connected the pure theory with some basic structures of
mathematics. In any case, the ``separation" referred to
above gave rise to a necessary and important internal
development of the subject. Even though there were good
relations and mutual admiration between the different
``camps", some people in group (i) were somewhat suspicious
of what they saw as overtly set-theoretic preoccupations in
Shelah's program and theory.

The last ten or fifteen years have seen a remarkable
unification or even re-unification of these differing trends
and emphases. One aspect is that the machinery and conceptual
framework of stability theory has been brought to bear on
the analysis of concrete structures in new ways. Related to
this is that various notions/dichotomies in stability theory
turn out to have meaning, not only for the general theory,
but for the (outside) mathematical world. (Actually I am here
talking not about stability theory per se but what one might
call ``generalized stability theory", the development of the
machinery of ``independence", dimension theory,
orthogonality, in model-theoretic contexts both broader than
and outside stable theories, such as simple theories and
o-minimal theories.) Going the other way, a kind
of sensitivity to the mathematical world, especially what
one may call a ``geometric sensibility" (complementing the
usual ``set-theoretic sensibility" characteristic of
logicians) has influenced the pure theory.

As a result of
these and earlier developments, model theory has assumed a
rather new role, complementing the classical
``foundations of mathematics". This is reflected in
Hrushovski's description of model theory as ``the geography
of tame mathematics". I give no definition of ``tame
mathematics" or ``tame structures". The real, $p$-adic, and
complex fields are tame. The ring of integers (and field of
rationals) are very nontame (or wild) as is any structure
which interprets them. Making sense of the tame/wild
borderline becomes a mathematical issue. Generalized
stability theory tends to rule out the interpretability of
wild structures.

Let me give a couple of examples of the unification referred
to above. The first
is the amazing journey from finite fields to the
``Independence Theorem" for simple theories. James Ax established the
decidability of the theory of finite fields in the 60's,
using among other things the Lang-Weil estimates for the
number of points on varieties over finite fields. In spite
of much work on pseudofinite fields and their
generalizations, pseudo-algebraically closed
fields, the connection with abstract
model-theoretic notions remained obscure. Shelah
\cite{Shelah} introduced simple theories in the late 70's as
theories without the ``tree property" (a certain
combinatorial property of formulas) generalizing stable
theories (theories without the ``order property"). His idea
was that the machinery of stability theory (such as forking)
might generalize to simple theories. Although Shelah made
several crucial insights, the situation remained
problematic, and the subject was not developed further until
the mid 90's. In the early 90's, Chatzidakis, van den Dries
and Macintyre, continuing Ax's work, gave a description of
definable sets in finite fields (and thus in the limit,
pseudofinite fields), associating to definable sets both
dimensions and measures, and asking several questions (such
as the status of ``imaginaries" in pseudofinite fields). I
remember receiving the preprint and leafing through it with
wonder late one afternoon in Notre Dame. Hrushovski
\cite{Hrushovski1} went further than I did.  He answered the
questions, in a more general context, theories of finite
$S_{1}$-rank, and proved the ``Independence Theorem" for
these theories, a result concerning the amalgamation of free
extensions of types. In the meantime Kim
\cite{Kim} had shown that the basic theory of forking does
indeed go through for Shelah's simple theories. Motivated by
Hrushovski's work (as well as Shelah's earlier work), this
Independence Theorem was proved for arbitrary simple
theories, and was moreover observed to be a {\em
characteristic} property of simple theories
\cite{Kim-Pillay}.
 
Another example is o-minimality. The notion of
$o$-minimality was developed both as an abstraction of the
properties of semialgebraic sets over the reals, and as an
analogue of strong minimality in the presence of a total
ordering (see \cite{vdDries}). In any case, if one allows the
notion of a total ordering as belonging to logic, the
classification of
$o$-minimal structures is an issue also of pure logic. A
theorem of Peterzil and Starchenko \cite{Peterzil-Starchenko}
recovers (expansions of) real
closed fields from o-minimality. This should be considered
as a foundational result in the new sense: from a notion of
pure logic one recovers model-theoretically (expansions of)
real algebraic geometry. Hrushovski and Zilber in an earlier
paper \cite{Hrushovski-Zilber} had already proved a similar
result for ``Zariski geometries", recovering algebraic
geometry.

Major results of the 90's were Wilkie's
proof of model-completeness (and o-minimality) of the real
field equipped with the exponential function \cite{Wilkie},
and Hrushovski's proof of the Mordell-Lang conjecture for
function fields in all characteristics \cite{Hrushovski2}.
Wilkie's ingenious proof made use of the general theory of
o-minimality. There is continuing work on finding richer
o-minimal expansions of the real field.
Hrushovski's work was informed by almost all
the accumulated results in stability theory, geometric
stability theory and stability-theoretic algebra
(differentially closed fields and separably closed fields).
 From this work and ongoing work by Hrushovski and others
(such as Scanlon) one sees that the
model-theoretic/stability-theoretic distinction between
linear (or modular) and nonlinear (nonmodular) behaviour of
definable sets has meaning in the world of geometry and
number theory.

The terms ``applied model theory" and ``applied
model-theorists" have been recently bandied around by
various people, to describe in a blanket fashion much
of the current work in model theory and its
practitioners.  I hope that the above discussion and
examples show that this is just wrong and moreover
completely misses the point. Although individuals may
choose to view themselves as ``applied", what is
specific to current developments is {\em not} a shifting of
attention to the external mathematical world, but the mutual
interaction between external and internal points of view,
and the corresponding enrichment of both.

There is now a reasonably coherent sense of what it means to
understand a structure: it means understanding the category
of definable sets (including quotients by definable
equivalence relations). Generalized stability theory gives a
host of concepts and tools which inform this analysis:
dimension theory (the assignment of meaningful ordinal-valued
dimensions to definable sets, invariant under definable
bijection), orthogonality, geometries, definable groups and
homogeneous spaces. As mentioned earlier, the contexts in
which such tools are applicable tend to rule out  G\"odel
undecidability phenomena. Interpretability is a key (even
characteristic) notion, and in a tautological sense the
business of ``pure" model theory becomes the classification
of first order theories up to bi-interpretability.

It is worth pointing out what some may consider paradoxical
in foundations, model theory  and the wild/tame distinction.
 From a classical foundational point of view the objects of
mathematics which can be most immediately grasped are the
``accessible domains" referred to in Sieg's talk (such as the
set of natural numbers equipped with all its arithmetic
operations). On top of these are
built the set-theoretically more complicated objects of
mathematics. In fact it is some of these {\em latter}
objects (such as locally compact fields), which, once
their set-theoretic genesis is forgotten, we have
direct access to, via quantifier-elimination and decidability
theorems. The accessible
domains, such as number fields and their absolute Galois
groups, although among the central objects in mathematics,
remain mysterious in many ways. It is typical in mathematics
to approach problems about these objects via tame objects
(such as via the Hasse principle and its obstructions).

There are many important current areas of research in and
around model theory which are not directly included in the
above discussion. The model theory of modules has been a
particularly active area. The Ziegler spectrum of a ring,
originating from  model-theoretic considerations
(positive-primitive formulas) is now a key notion and tool
in the representation theory of rings. In this case too, the
stability-theoretic perspective has been important. Work on
generalized quantifiers and infinitary logic continues,
especially in the context of ``nonstructure theorems" by
Shelah and his collaborators. The subject ``finite model
theory", the study of definable classes of finite structures
and definability in finite structures has also been rather
active, with connections to computer science and complexity.
In this context first order definability is often the wrong
notion to consider and either fragments (such as first
order logic with finitely many variables) or other logics are
more appropriate. Even in the context of first order
definability on infinite structures, nonfirst order
considerations naturally arise, for example when one wants
to consider type-definable sets and even their quotients by
type-definable equivalence relations as structures in their
own right. Although nonstandard analysis has long ago become
a separate subject, model theory has been enriched by the
development (by Keisler \cite{Keisler}, Henson and others)
of appropriate logics and tools for dealing with metric
spaces, Banach spaces and the like.

\FOLsubsection{The future.}
I will not try to predict developments but will limit myself
to discussing a few ``themes" (and problems) which are mostly
related to the current developments discussed above. This is
of course both limited and influenced by my own
knowledge and preoccupations.

\vspace{2mm}
{\em Foundations of model theory.}
What is the right language and level of generality for
model theory? The traditional framework of one-sorted
structures and their point-sets has long been recognized as
being rather restrictive. The actual practice of
model-theorists is somewhat more in line with
points of view from categorical logic. Moreover a degree of
flexibility is required to deal with various
natural elaborations of and variants of first order
definability.

\vspace{2mm}
{\em Classification of uncountably categorical and related
structures, up to bi-interpretability.}
This is a rather strong formulation of Zilber's program.
In this form it will probably never be
accomplished, but it subsumes an enormous amount of work in
model theory: the geometry of strongly minimal
sets, the mathematics around Hrushovski's
amalgamation/fusion techniques, the structure of simple
noncommutative groups of finite Morley rank (Cherlin's
conjecture) and the theory of covers. Included in ``related
structures" are the structures of finite $SU$-rank, say,
where much of the geometric theory has still to be
developed.

Interpreted more loosely we could include here
ongoing work in  stability, its generalizations (such as
simple theories), and the classification of first order
theories.

\vspace{2mm}
{\em Model theory and analysis/geometry.}
It is hoped that the second part of Hilbert's 16th problem
(uniform bounds on the number of limit cycles of polynomial
planar vector fields) can be approached by finding suitably
rich $o$-minimal expansions of the real field.

The understanding of complex exponentiation is a major
challenge, in particular Zilber's conjecture that the
complex field equipped with the exponential function is
``tame" modulo countable definable sets.

Bimeromorphic geometry is concerned with the classification
of compact complex manifolds up to bimeromorphic
equivalence. There is a hope that geometric
stability-theoretic methods would yield nontrivial results
here, although maybe it is too early to tell. There are
intriguing connections with ``$o$-minimal
complex analysis".

I also include here further development of model theoretic
techniques and notions appropriate for metric spaces, Banach
spaces and function spaces, as well as applications.

\vspace{2mm}
{\em Model theory and number theory.}
The kind of model-theoretic methods discussed in this article
have not yet penetrated the central
problems concerning rational points (namely over number
fields) of varieties. This is a major challenge, and any
progress would have to incorporate  arithmetic features such
as heights into associated model-theoretic structures.

On the other hand, various theorems about rational points
(such as Mordell-Lang over number fields) have {\em
equivalent} model-theoretic statements (although not as yet
model-theoretic proofs). In fact we have a new twist on the
notion  ``fragments of arithmetic": Fix a variety $V$
defined over ${\bf Q}$ and let $M_{V}$ be the structure
$({\bf C},+,\cdot, V({\bf Q}))$ (so we adjoin a predicate
for the rational points of $V$ to the complex field). Is it
the case that $M_{V}$ is either stable or undecidable? What
are the possible Turing degrees of such structures? Are
these questions settled by the Lang conjectures on varieties
of general type?

\vspace{2mm}
{\em Model theory and differential equations.}
I mean here the algebraic theory of differential equations
and the structure of solution sets. Concerning ordinary
differential equations, the fine structure of definable sets
of finite Morley rank in differentially closed fields is
relevant. A challenge is to extend finiteness theorems for
equations of order $1$ to the higher order case.  For partial
differential equations, infinite-dimensional sets (i.e.
definable sets of infinite Morley rank in the appropriate
structures) come into the picture, and are hardly understood
at all model-theoretically.

Another important problem is to identify
and axiomatize the  universal domains appropriate for the
kind of ``asymptotic differential algebra" embodied in Hardy
fields.

Finally, one would hope for model-theoretic methods
(especially those discussed in this article) to be relevant
to Grothendiek's conjecture in the arithmetic of linear
differential equations.

\vspace{2mm}
{\em Finite and pseudofinite structures.}
I am referring here to the (first order) model-theoretic
study of infinite limits (in various senses) of finite
structures, and the light this sheds on uniformities in
families of finite structures. (So this is not exactly the
same as so-called finite model theory.) The work on
smoothly approximable structures (Lachlan, Cherlin,
Hrushovski \cite{Cherlin-Hrushovski}, Kantor, Liebeck,
Macpherson and others) as well as work on pseudofinite
groups and fields falls under this rubric. The content and
implications of {\em pseudofiniteness} (being an
ultraproduct of finite structures) is an important issue.

\vspace{2mm}
{\em Hilbert's 10th problem over ${\bf Q}$.}
This is very much related to the number theory discussion
above. The problem is whether there is an effective way of
deciding, given a finite system of polynomials in several
variables with rational coefficients, whether or not this
system has a solution all of whose coordinates are rational
numbers. Formulated logically it is the problem of the
decidability of the existential theory of $({\bf
Q},+,\cdot)$. (The full theory is undecidable.) Formulated
geometrically it is the problem of deciding the existence of
rational points on varieties defined over ${\bf Q}$. A
negative solution would follow from being able to
existentially define the ring ${\bf Z}$ in the field
${\bf Q}$. (A possible obstruction to this is a certain
conjecture of Barry Mazur on the {\em topology} of rational
points of varieties.) It is rather interesting that number
theorists appear to favour a positive solution to the main
problem. In fact in the case of curves ($1$-dimensional
varieties), it has been conjectured that one can even
{\em compute} the set of rational points.

\vspace{2mm}
{\em Vaught's conjecture.}
Vaught's conjecture for first order countable theories
remains open: a first order countable theory has either at
most $\omega$ or exactly $2^{\omega}$ countable models. One
hopes for a renewal of the ``approach from below" started
by Shelah for the $\omega$-stable case and
continued by Buechler and Newelski for the superstable of
finite rank case. It would be nice to see also an approach
from above. In the more general context of
$L_{\omega_{1},\omega}$ theories, it is a special case of
the Topological Vaught conjecture from descriptive set
theory.

\vspace{2mm}
{\em Logic and mathematics.}
One theme in the discussion following the panel presentation
was: how can logic increase its prestige within mathematics
and how does one go about making a ``splash" which
mathematicians will take notice of? My feeling is that this
is the wrong sort of question. If one wants some kind of
meaningful interaction with other parts of mathematics, it
is the conviction that this is a worthwhile
intellectual enterprise, rather than the desire to make a
``splash", which is crucial. This conviction amounts
essentially to a belief in the unity of mathematics. There
has been much discussion of this ``unity of mathematics" in
recent times, often in connection with deep conjectures
relating arithmetic, geometry, analysis, representation
theory etc. One feels moreover that logicians, especially in
the light of their foundational concerns, should have some
level of engagement with these issues and conjectures.
There is another
sociological aspect. In so far as logicians live and operate
within mathematics departments there is a need to talk to
and interact with the people around them. So the issue is
that of a sensitivity to mathematics and educating our
graduate students accordingly. I believe that with such a
sensitivity, interactions and ``splashes" will take care of
themselves, and our subject, or rather its various branches,
may end up being transformed in the process.

{\em Acknowledgement.} Thanks to Lou van den Dries for his
comments on an earlier draft.

\section{Set Theory}
\sectionAuthor{Alexander Kechris}

A)  I will present here some speculations on future directions
in set theory.  Modern set theory is a vast and very diverse
subject, so it is obvious that in a short time I cannot possibly
cover all important aspects of research in this field.  I will also
concentrate on discussing, in fairly broad terms, general programs
and trends, as opposed to specific problems, with some obvious
exceptions.

B) For the purposes of this presentation, it will be convenient
for me to distinguish two aspects of research in set theory:

The first, which I will call {\it internal} or {\it foundational},
is concerned with the understanding and clarification of the
basic concepts of set theory itself, and aims at providing a
foundation for a comprehensive and satisfactory theory of sets.
Since the time of Cantor, set theory has been continuously
evolving towards that goal and this trend will undoubtedly
continue in the future.

The second aspect, which I will call {\it external} or {\it
interactive}, deals with the connections of set theory with
other areas of mathematics.  This includes the use of set
theoretic concepts, methods, and results in establishing the
foundations or helping the development of other mathematical
disciplines as well as the application of set theoretic techniques
in the solution of specific problems in such areas.

Of course, these two aspects, internal and external, are often
closely interrelated.

C)  Also, following a well-established tradition going back to
Cantor, it will be useful to subdivide the theory of
sets into (i) the {\it theory of the continuum} or {\it theory
of pointsets}, i.e., the study of sets and functions on the reals,
complex numbers, Euclidean spaces or, more generally, Polish
(complete separable metric) spaces, and (ii) the {\it general
set theory} of arbitrary sets and cardinals.

An important further distinction in the theory of the continuum
was introduced in the early 20th Century by the French, Russian,
and Polish analysts, who laid the foundations of {\it descriptive
set theory} or {\it definability theory of the continuum}, which is
the study of definable (e.g., Borel, projective, etc.) sets and
functions on Polish spaces.  So we can subdivide the theory of the
continuum into {\it descriptive set theory} and the {\it theory
of arbitrary pointsets}.  For example, a question such as the
measurability of the projective sets belongs to the first part but
the Continuum Hypothesis (CH) or the study of cardinal characteristics of the continuum belongs to the second.

Again all these aspects of set theory are closely interrelated.
With these classifications in mind, I will now discuss some
prospects for research in set theory.

\FOLsubsection{Descriptive set theory.}

A)  Work in the last 30 years or so has resulted in a resolution
of the foundational (internal) issues facing descriptive set
theory.  There is now a very satisfactory and comprehensive
foundation for the theory of definable sets and functions on
Polish spaces, based on the principle of {\it Definable
Determinacy} (see Kechris~\cite{ke1995}, Moschovakis~\cite{ynm1980}).  This
theory also fits beautifully within the framework of global
set theory as currently developed through the theory of large
cardinals.  The determinacy principle is in fact ``equivalent",
in an appropriate sense, to the existence of certain types of
large cardinals (see Martin-Steel~\cite{ms1989}, Woodin~\cite{w1988}).
Moreover the structure theory of definable sets in Polish spaces,
that determinacy unveils, has a very close and deep
relationship with the unfolding inner model theory of large
cardinals, an example of which was so vividly illustrated in 
Itay Neeman's talk in this conference.

B) Thus the foundational aspects of descriptive set theory are by 
and large settled now.  Research in this area is now increasingly
preoccupied with external issues. These consist of applying the
ideas, methodology, and results of descriptive set theory, both
in its classical and modern manifestations, to other areas of
mathematics, while at the same time developing new directions
in the theory itself which are motivated by such interactions.
Interestingly, this also leads to the use of sophisticated methods
and results from other areas of mathematics in the solution of
purely set theoretic problems in descriptive set theory.

Intriguing such connections have been discovered during the last
15 years or so in areas such as classical real analysis, harmonic
analysis, Banach space theory, and ergodic theory (see, for
example, Foreman et al.~\cite{fklw}, 
Kahane-Salem~\cite{ks1994}, Kechris-Louveau~\cite{kl1989}).
More recently, a very promising new area, that
is now very actively investigated, deals with the development of
a {\it theory of complexity of classification problems in
mathematics}, a classification problem being the question of
cataloging a class of mathematical objects up to some notion of
equivalence by invariants, and the closely related theory of
{\it descriptive dynamics}, i.e., the theory of definable actions
of Polish groups on Polish spaces (see Becker-Kechris~\cite{bk1996},
Hjorth~\cite{hj2000},
Kechris~\cite{ke2001}).  This work brings
descriptive set theory into contact with current developments
in various areas of mathematics such as dynamical systems,
including ergodic theory and topological dynamics, the theory
of topological groups and their representations, operator
algebras, abelian and combinatorial group theory, etc.  Moreover,
it provides new insights in the traditional relationships of descriptive set theory with other areas of mathematical logic, as, for example, with recursion
theory, concerning the global structure of Turing degrees (see
P. Cholak et al.~\cite{cll2000}),
or with model theory, through the Topological Vaught Conjecture
and the general study of
the isomorphism relation on countable structures.

Moving beyond descriptive set theory, I will concentrate on
two other major aspects:  the {\it theory of large cardinals}
and the {\it theory of small cardinals}.

\FOLsubsection{The theory of large cardinals.}

A) The goal of the theory of large cardinals is to understand the
higher reaches of infinity and their effect on its lower levels.
Another important aspect here is the use of large cardinal
principles as a global scale for calibrating the consistency
strength of extensions of classical ZFC set theory.

Most of the effort in this area today is going towards the
internal or foundational aspects of this theory, where a vigorous
and far reaching program is actively pursued, dealing with the
development of canonical inner models for large cardinals and
the detailed analysis of their structure, as well as their
relationship with descriptive set theory (see L\"owe-Steel~\cite{ls1999},
Steel~\cite{sinfty}).  This program is by no means complete
yet, and it will undoubtedly be one of the main topics of set
theoretic research in the foreseeable future.  It is also
closely interrelated to many other important directions of 
research in set theory, including infinite combinatorics and
the development of forcing techniques (see the forthcoming
Foreman et al.~\cite{fkm}).

B)  Simultaneously with the pursuit of the foundational goals,
there have been several interesting external developments here
as well.  There is of course a long tradition of application
of set theoretic techniques, involving for example, forcing,
infinite combinatorics as well as large cardinals, to many areas
of abstract algebra, functional analysis, measure theory and
general topology, for instance in obtaining significant
independence and consistency results, as for example in the
Whitehead Problem (Shelah; see Shelah~\cite{ss1974}), the Kaplansky
Conjecture (Dales, Esterle, Solovay, Woodin; see Dales-Woodin~\cite{dw1987}),
or the S- and L- space problems (see, for example, Todorcevic~\cite{st1989}),
and this will of course continue in the future.
More recently, large cardinal theory is finding its way into more
concrete situations.  H. Friedman (see, for example, Friedman~\cite{f1998})
applies combinatorics of large cardinals to obtain new
combinatorial principles for finite sets.  Moreover he shows that
these principles require, in an appropriate sense, these large
cardinal hypotheses.  Another interesting direction relates the
structure of elementary embeddings associated with large cardinals
to that of self-distributive algebras and braid groups, through
work of Laver, Dehornoy, and others (see Dehornoy~\cite{dinfty}).

\FOLsubsection{The theory of small cardinals.}

A)  In this context, I include both the study of arbitrary
pointsets, in particular problems such as the CH, as well as the
theory of the ``small" alephs $\aleph_1,\ \aleph_2$,...

Here the foundational situation is far from clear.  
The theory of large cardinals has many important
implications here, in particular in terms of consistency
and independence results (see, for instance, 
Foreman-Magidor-Shelah~\cite{fms1988}).
However, it is well-known that in its present form,
which is largely immune to forcing constructions, it does
not resolve key issues such as the CH.  It is clear that
a satisfactory and comprehensive theory of small cardinals
needs to be developed, within which we can hope to achieve
the resolution of this basic set theoretic problem and
related questions.

B)  A promising new approach along these lines has been
recently initiated by Woodin (see Woodin~\cite{w1999}), which
aims at developing a theory that leads to a ``complete"
understanding of the definability structure of the power
set of $\omega_1,\ P(\omega_1)$, which will parallel
the ``complete" understanding of the definability structure
of the power set of $\omega ,\ P(\omega )$, based on the
principle of definable determinacy.  Towards developing
such a theory, Woodin proposes a new principle, concerning
the definability structure of (an enriched form of) 
$P(\omega_1)$, which implies the failure of the CH, in
fact it gives the answer $2^{\aleph_0}=\aleph_2$ for the
value of the cardinality of the continuum.  It is of course too early yet to
know the full effect of this theory, and whether it will
be eventually viewed as the ``right" theory for the 
definability structure of $P(\omega_1)$, finally leading
to a satisfactory resolution of the CH.  This will require
a much more detailed development of the theory than is
presently available, and should be the focus of extensive research in the future.  Even if this turns out to
be successful, further questions concerning the theory of
arbitrary pointsets and the structure of small cardinals
would need the development of a theory of $P(\omega_2),
P(\omega_3)\dots$, for which no hints are available at this
stage.

{\it Acknowledgment.} I am grateful to Yiannis Moschovakis, Richard Shore, John Steel, and Hugh Woodin for their comments on an earlier draft of this section.

\end{document}